\documentclass[a4paper,12pt]{article}
\usepackage{epsfig}

\date{\today}

\usepackage{amsmath}
\usepackage{amssymb}
\usepackage{float}
\usepackage[hang,nooneline,scriptsize]{subfigure}
\begin{document}

\title{Minimal boson stars in 5 dimensions:\\ classical instability and existence of ergoregions}
\vspace{1.5truecm}
\author{
{\bf Yves Brihaye$^{1,}$\footnote{yves.brihaye@umons.ac.be} \ and Betti Hartmann$^{2,}$}\footnote{bhartmann@ifsc.usp.br}\\[10pt]
$^1$ Physique-Math\'ematique, Universit\'e de
Mons, 7000 Mons, Belgium\\
\hspace{0.1cm}\\
$^2$ Instituto de F\'isica de S\~ao Carlos, Universidade de S\~ao Paulo,\\
 Caixa Postal 369, CEP 13560-970, S\~ao Carlos (SP), Brazil
}

\vspace{1.5truecm}

\date{\today}
\setlength{\footnotesep}{0.5\footnotesep}
\newcommand{\dd}{\mbox{d}}
\newcommand{\tr}{\mbox{tr}}
\newcommand{\la}{\lambda}
\newcommand{\ka}{\kappa}
\newcommand{\f}{\phi}
\newcommand{\vf}{\varphi}
\newcommand{\F}{\Phi}
\newcommand{\al}{\alpha}
\newcommand{\ga}{\gamma}
\newcommand{\de}{\delta}
\newcommand{\si}{\sigma}
\newcommand{\bomega}{\mbox{\boldmath $\omega$}}
\newcommand{\bsi}{\mbox{\boldmath $\sigma$}}
\newcommand{\bchi}{\mbox{\boldmath $\chi$}}
\newcommand{\bal}{\mbox{\boldmath $\alpha$}}
\newcommand{\bpsi}{\mbox{\boldmath $\psi$}}
\newcommand{\brho}{\mbox{\boldmath $\varrho$}}
\newcommand{\beps}{\mbox{\boldmath $\varepsilon$}}
\newcommand{\bxi}{\mbox{\boldmath $\xi$}}
\newcommand{\bbeta}{\mbox{\boldmath $\beta$}}
\newcommand{\ee}{\end{equation}}
\newcommand{\eea}{\end{eqnarray}}
\newcommand{\be}{\begin{equation}}
\newcommand{\bea}{\begin{eqnarray}}

\newcommand{\ii}{\mbox{i}}
\newcommand{\e}{\mbox{e}}
\newcommand{\pa}{\partial}
\newcommand{\Om}{\Omega}
\newcommand{\vep}{\varepsilon}
\newcommand{\bfph}{{\bf \phi}}
\newcommand{\lm}{\lambda}
\def\theequation{\arabic{equation}}
\renewcommand{\thefootnote}{\fnsymbol{footnote}}
\newcommand{\re}[1]{(\ref{#1})}
\newcommand{\R}{{\rm I \hspace{-0.52ex} R}}
\newcommand{\N}{{\sf N\hspace*{-1.0ex}\rule{0.15ex}%
{1.3ex}\hspace*{1.0ex}}}
\newcommand{\Q}{{\sf Q\hspace*{-1.1ex}\rule{0.15ex}%
{1.5ex}\hspace*{1.1ex}}}
\newcommand{\C}{{\sf C\hspace*{-0.9ex}\rule{0.15ex}%
{1.3ex}\hspace*{0.9ex}}}
\newcommand{\eins}{1\hspace{-0.56ex}{\rm I}}
\renewcommand{\thefootnote}{\arabic{footnote}}
 \maketitle
\begin{abstract} 
We show that {\it minimal boson stars}, i.e. boson stars made out of scalar fields without self-interaction, are always {\it classically unstable} in 5 space-time dimensions. 
This is true for the non-rotating as well
as rotating case with two equal angular momenta and in both Einstein and Gauss-Bonnet gravity, respectively, and contrasts with the 4-dimensional case, where classically
stable minimal boson stars exist.
We also discuss the appearance of ergoregions for rotating boson stars with two equal angular momenta. While rotating
black holes typically possess an ergoregion, rotating compact objects without horizons such as boson stars
have ergoregions only in a limited range of the parameter space. In this paper, we show for which values of the parameters
these ergoregions appear and compare this with the case of standard Einstein gravity. We also point out that the interplay between
Gauss-Bonnet gravity and rotation puts constraints on the behaviour of the space-time close to the rotation axis. 
\end{abstract}

\section{Introduction}
Non-topological solitons are distinct from topological solitons \cite{ms} in the
sense that while the latter possess a charge that is of topological origin, the former
possess a conserved Noether charge that arises from a continuous symmetry inherent in
the model. The best known example of a non-topological soliton is the $Q$-ball,
where the $Q$ refers to the conserved Noether charge \cite{fls,lp,coleman}.
The model contains a complex scalar field and possesses a global U(1) symmetry.
Localized objects in the soliton-sense are possible if a self-interaction
potential for the scalar field is introduced. This allows for a subtle interplay 
between quantum mechanical principles and the scalar field self-interaction. 
In \cite{vw,kk1,kk2} non-rotating and rotating $Q$-balls in $(3+1)$ space-time dimensions have been constructed using a non-renormalizable scalar field potential of 6th order in the scalar field, while in supersymmetric 
extensions of the Standard Model $Q$-balls also exist for more complicated scalar field potentials \cite{kusenko,cr,ct}. These have been discussed in detail for a 
scalar field potential of exponential form arising in gauge-mediated supersymmetry breaking \cite{cr,ct,Hartmann:2012gw}.

The self-gravitating counterparts of $Q$-balls, so-called {\it boson stars} have also been discussed extensively 
\cite{kaup,Ruffini:1969qy,misch,flp,jetzler,new1,new2,Liddle:1993ha}. In this
case, it is sufficient
to introduce a mass term for the scalar field, while self-interaction of the scalar field is not necessary for boson stars to exist. Following the literature, we will refer to these boson stars made out of a massive scalar field
without self-interaction as to {\it minimal boson stars}. These have been
studied for the first time in \cite{kaup}. In most studies of
non-rotating and rotating boson stars, however, solutions with a flat space-time limit have been discussed, such that the scalar field of the boson
star is always self-interacting. In 4 space-time dimensions, these solutions have been
studied in great detail in \cite{Kleihaus:2005me,Kleihaus:2007vk} using a 6th order scalar field potential and an exponential potential in \cite{Hartmann:2012gw}, respectively.

While boson stars in 4 space-time dimensions are interesting from an astrophysical perspective \cite{schunck_liddle} as well as when considering
the discovery of a fundamental scalar field in nature \cite{cern_lhc}, they can also be used to gain more insight into the fundamental properties of the gravitational interaction. Most current suggestions for a Quantum Theory of gravity
require the existence of extra dimensions. Now, one would additionally expect Quantum gravity effects to become important for strong gravitational fields.
The gravitational fields of boson stars can be strong, but the space-time does not possess horizons. 
It is hence interesting to study these strongly self-gravitating systems and compare their properties with those of black holes.
Non-rotating boson stars 
in $d$-dimensional Einstein gravity with $d=3,5,6,7$ have been studied in
\cite{Hartmann:2010pm} using an exponential scalar field potential and it was noticed
that the critical behaviour of the solutions depends on the number of space-time dimensions. Moreover, non-rotating boson stars in 5-dimensional
Einstein-Gauss-Bonnet (EGB) gravity have been considered in \cite{Hartmann:2013tca} and it
was shown that the qualitative features of the solutions change when the Gauss-Bonnet term dominates the gravitational interaction.

The study of boson stars in higher dimensions is also interesting from another point of view: while rotating objects in 4 space-time dimensions possess only one angular momentum, they can possess more than one angular momentum in more than 4 dimensions
due to the existence of additional (orthogonal) planes of rotation. In 5 space-time dimensions, e.g., two orthogonal planes of rotation exist and hence
rotating objects can possess two angular momenta. In the case of equal angular momenta the symmetry can be enhanced such that the metric functions depend on the radial coordinate only. Rotating boson stars in 5-dimensional space-time with two equal angular momenta
were discussed for the first time in \cite{Hartmann:2010pm} using a 6th order scalar field potential. It was shown that the sum of the angular momenta is proportional to the Noether
charge in this case. 
This study has been extended to include the Gauss-Bonnet (GB) interaction in \cite{Brihaye:2013zha,Henderson:2014dwa} and it was shown that rotating
boson stars in EGB gravity do not exist when the Gauss-Bonnet interaction
dominates the gravitational interaction. Using a perturbative expansion it was
shown that the solutions cease to exist in this case \cite{Henderson:2014dwa}. In this paper we point out that this can
be traced back to the interplay between the scalar field function and the metric
functions close to the origin.

The classical stability of both $Q$-balls and boson stars is of crucial importance. Considering $Q$ to denote the number of scalar particles of which the
boson star is made off, the total mass $M$ can be compared with the mass of $Q$ scalar bosons with mass $m$.
For $M < m Q$ we would expect the boson star to be {\it classically stable} in the sense that the kinetic energy of $Q$ scalar bosons
can be balanced by the gravitational energy of the system.
The first detailed study of this type was done for minimal boson stars in 4 space-time dimensions \cite{Lee:1988av} and it was shown that minimal boson stars in 4 space-time dimensions can be classically stable. In all studies including self-interaction 
of the scalar field it was found that stable as well as unstable boson star solutions
exist (see \cite{Kleihaus:2011sx} for a recent study). This is true for non-rotating as well as rotating boson stars in 4 and more dimensions. In general, it was found that the increase of the gravitational coupling leads to a decrease in the relative extent of the
classically stable branch with respect to the classically unstable branch. For sufficiently large gravitational interaction the solutions
are always classically unstable. 

One of the objectives of this paper is to point out that minimal boson stars in 5 space-time dimensions are always 
classically unstable - both for the non-rotating and the rotating
case as well as for Einstein gravity and Gauss-Bonnet gravity, respectively. This is remarkable since in 4 space-time dimensions stable minimal boson stars exist
and thus demonstrates that the number of space-time dimensions can influence the stability. We point out below that this is related to the $1/r^2$ fall-off of the
gravitational potential as compared to the $1/r$ fall-off in 4 dimensions. 
Let us remark that rotating minimal boson stars in 5-dimensional Einstein gravity have been considered in the context of
rotating black holes carrying scalar hair, where they appear as limiting solutions \cite{brihaye_radu_herdeiro}.

Another objective of this paper is the discussion of the ergoregions of the rotating
boson stars. In general, rotating objects can possess an ergoregion in which the
asymptotically time-like Killing vector becomes space-like. As pointed out in \cite{Press:1972zz, Cardoso:2004nk} this can lead to
a superradiant instability because infalling
bosonic waves are amplified when reflected. 
For boson stars it has been argued that the appearance of ergoregions leads
to an instability, the so-called {\it ergoregion instability} \cite{Cardoso:2007az}. 
In contrast to black holes, the space-time is globally regular and possesses no horizons, such that scattered waves
that can escape to infinity carry away energy and by such a process can destabilize the star.

In this paper we demonstrate that rotating boson stars in 5 space-time dimensions
can also possess ergoregions, but only for sufficiently large increase of the
scalar field function close to the origin. We also show that the GB interaction
changes the features of these ergoregions only marginally. 

Our paper is organized as follows: in Section 2 we introduce the model and give the Ansatz. In Section 3, we discuss our numerical results. This includes the
discussion of the interplay between rotation and the Gauss-Bonnet interaction, of the (in)stability of minimal boson stars as well as of the appearance of ergoregions, respectively. Section 4 contains our conclusions and outlook.

\section{The model}
In this paper we study Einstein-Gauss-Bonnet gravity in 5 space-time dimensions coupled to a complex scalar field that can possess self-interaction.
The action of this model reads (assuming natural units such that $\hbar=c=1$)
\be
\label{egbbs}
   S = \frac{1}{16 \pi G_{\rm 5}} \int d^5 x \left[ R  + \frac{\alpha}{2} {\cal L}_{\rm GB} 
   - 16 \pi G_{\rm 5} \left( \partial_M \Pi^{\dagger} \partial^M \Pi + m^2 \Piˆ^{\dagger} \Pi + V_{\rm si}(\Pi^{\dagger}\Pi) \right) \right]      \ ,   
\ee
where $R$ is the Ricci scalar, $\Pi$ is a complex scalar doublet with mass $m$ and self-interaction potential $V_{\rm si}(\Pi^{\dagger}\Pi)$,  $\alpha$ denotes the Gauss-Bonnet coupling constant
and the Lagrangian density of Gauss-Bonnet gravity reads
\be
        {\cal L}_{\rm GB} = R^{MNKL} R_{MNKL} - 4 R^{MN} R_{MN} + R^2 \ \ , 
\ee
with $M,N,K,L \in \{0,1,2,3,4 \}$.  Newton's constant $G_{\rm 5}$ in 5 dimensions is related to the Planck mass $M_{\rm Pl,5}$ and Planck length $l_{\rm Pl,5}$ in 5 dimensions, respectively, 
by $G_5=M_{\rm Pl,5}^{-3}=l_{\rm Pl,5}^3$, while the Gauss-Bonnet coupling $\alpha$ has the dimension of a $[{\rm length}]^2$. If we would treat the action above as a low energy effective action of String Theory, $\alpha$ would fulfill $\alpha \sim l_{\rm Pl,5}^2$.

The equations of motion then read
\begin{equation}
 G_{MN} +  \frac{\alpha}{2} H_{MN}=8\pi G_5 T_{MN} \ ,  M,N=0,1,2,3,4 \ ,
\end{equation}
where $H_{MN}$ is given by
\begin{eqnarray}
 H_{MN}&=& 2\left(R_{M ABC}R_{N}^{ABC} - 2 R_{M A N B}R^{AB} - 2 R_{M A}R^{A}_{N} + 
R R_{MN}\right)  \nonumber \\
&-& \frac{1}{2} g_{MN} \left(R^2 - 4 R_{AB}R^{AB} + R_{ABCD} R^{ABCD}\right) \ , \ \ A,B,C=0,1,2,3,4 \ , 
\end{eqnarray}
and $T_{MN}$ denotes the energy-momentum tensor of the scalar field
\begin{eqnarray}
\label{em}
T_{MN}&=& g_{MN} {\cal L} - 2\frac{\partial {\cal L}}{\partial g^{MN}}
= \partial_{M} \Pi^{\dagger} \partial_{N} \Pi + \partial_{N}\Pi^{\dagger} \partial_{M} \Pi \nonumber \\
&-&g_{MN} \left[\frac{1}{2} g^{KL} 
\left(\partial_{K} \Pi^{\dagger} \partial_{L} \Pi +
\partial_{L} \Pi^{\dagger} \partial_{K} \Pi\right) +m^2 \Pi^{\dagger}\Pi+V_{\rm si}(\Pi^{\dagger}\Pi)\right]   \ .
\end{eqnarray}
The scalar field equation is given by the Klein-Gordon equation
\begin{equation}
\label{KG}
 \left(\square - m^2 - \frac{\partial V_{\rm si}}{\partial \vert\Pi\vert^2} \right)\Pi=0 \ \   \ \ .
\end{equation}
In this paper, we are interested in two cases: (a) boson stars composed of a massive
scalar field without self-interaction, i.e. $V_{\rm si}\equiv 0$ (following the literature we will refer to these boson stars 
in the following as {\it minimal boson stars}) and (b) boson stars with a self-interaction of the form
\begin{equation}
\label{pot_self}
V_{\rm si}(\vert\Pi\vert^2)= m^2 \eta^2 \sum\limits_{n=2}^{\infty} (-1)^{n+1}  \left(\frac{\vert\Pi\vert}{\eta}\right)^{2n}   \ ,
\end{equation}
where $\eta$ is a (dimensionful) coupling constant. This self-interaction potential appears in gauge-mediated supersymmetry
breaking with breaking scale $\eta$ \cite{cr,ct}. We will concentrate on case (a),
but also present some results for case (b). 

\subsection{Ansatz}

In principle, localized objects in 5-dimensional space-time
can possess two independent angular momenta associated to the two orthogonal planes of rotation. 
If one restricts to  the case of equal angular momenta the symmetry of the object is enhanced and the
Ansatz for the metric reads \cite{Hartmann:2010pm}
\begin{eqnarray}
\label{metric}
ds^2 & = & -b(r) dt^2 + \frac{1}{f(r)} dr^2 + g(r) d\theta^2 + h(r)\left[\sin^2\theta \left(d\varphi_1 - 
W(r) dt\right)^2 + \cos^2\theta\left(d\varphi_2 -W(r)dt\right)^2\right] \nonumber \\
&+& 
\left(g(r)-h(r)\right) \sin^2\theta \cos^2\theta (d\varphi_1 - d\varphi_2)^2 \ ,
\end{eqnarray}
where $\theta\in [0:\pi[$, while $\varphi_k\in [0:2\pi[$, $k=1,2$. 
The corresponding space-time  possesses two rotation planes at $\theta=0$ and $\theta=\pi/2$ and the isometry
group is $\mathbb{R}\times U(2)$. 
The metric (\ref{metric}) still leaves the diffeomorphisms related to the definition of the radial variable $r$ unfixed. This can be fixed
by choosing $g(r)=r^2$ and we will employ this choice in the following. 

In order to construct rotating boson stars in 5 dimensions, the following Ansatz for the complex scalar doublet was first introduced in \cite{Hartmann:2010pm} and reads
\be
          \Pi(t,r,\theta,\varphi_1,\varphi_2) = \phi(r) e^{i \omega t} \hat{\Pi}(\theta,\varphi_1,\varphi_2)  \ ,
\ee
where $\hat \Pi$ is a doublet of unit length that depends on the angular coordinates only and is chosen such that
\be
\label{hatphi_rot}
\hat \Pi = \left(\sin \theta e^{i \varphi_1},\cos \theta e^{i \varphi_2}  \right)^t \ .
\ee  

Note that the case of non-rotating Gauss-Bonnet boson stars, which has been studied in \cite{Hartmann:2013tca}, corresponds
to the choice $\hat \Pi = (1,0)^t$. In this case, the equations lead to $W(r)\equiv 0$ and $h(r)=r^2$. 

While the metric (\ref{metric})
has three commuting Killing vector fields $\partial_t$, $\partial_{\varphi_1}$, $\partial_{\varphi_2}$ \cite{Hartmann:2010pm}, 
the scalar doublet of the rotating solution with (\ref{hatphi_rot}) is only invariant 
under one possible combination of these vectors, namely under
$\partial_t - \omega\left(\partial_{\varphi_1} + \partial_{\varphi_2}\right)$.

\subsection{Physical quantities}
The action possesses a global $U(1)$ symmetry which leads to the existence of
a globally conserved Noether charge $Q$ which reads \cite{Hartmann:2010pm}
\be
\label{Qeq}
Q= -\int \sqrt{-g} j^t  \ {\rm d}^4 x \ ,
\ee
where $j^t$ corrsponds to the $t$-component of the locally conserved Noether current $
j^{\mu} = -i\left(\Pi^{\dagger} \partial^{\mu} \Pi - \partial^{\mu} \Pi^{\dagger} \Pi\right)$. Inserting the specific Ansatz into (\ref{Qeq}) we
find in our choice of metric:
\begin{equation}
Q=4\pi^2 \int \sqrt{\frac{bh}{f}}\frac{r^2}{b} \left(\omega + W\right)  \phi^2 {\rm d} r  \ .
\end{equation}
Using the Komar expressions to evaluate the angular momenta $J_1=J_2\equiv J$ it was realized that $Q=\vert J_1\vert + \vert J_2\vert = 2\vert J\vert$ \cite{Hartmann:2010pm}.

The mass of the solutions can be given by the relevant Komar expression (see e.g. \cite{Hartmann:2010pm}) and is proprtional
to the prefactor in the $1/r^2$ fall-off of the metric function $b(r)$ at infinity (see e.g. \cite{Brihaye:2010wx}).

The Ricci scalar, which describes the local scalar curvature of the space-time reads
\begin{eqnarray}
R &=&  \frac{8}{r^2} -\frac{2h}{r^4} + f\left(-\frac{h''}{h} + \frac{h'^2}{2h^2} - \frac{2h'}{rh} -\frac{4}{r^2} +  2  - \frac{b'h'}{2bh}-
      \frac{2b'}{br} - \frac{b''}{b}  + \frac{h W'^2}{2b} + \frac{b'^2}{2b^2}\right) \nonumber \\
      &-& f'\left( 
 \frac{h'}{2h} +  \frac{2}{r} + \frac{b'}{2b} \right) \ ,
 \end{eqnarray}
 where the prime denotes the derivative with respect to $r$.
 
Recall that in pure Einstein gravity, i.e. for $\alpha=0$ we have $3R=-16\pi G_{\rm 5} T$, where $T$ denotes the trace of the energy-momentum tensor. 

\section{Results}
In the following, we will discuss our numerical results that we have obtained by solving the coupled system of ordinary differential equations
by employing a Newton-Raphson adaptive grid iteration scheme \cite{colsys}. 

We apply the following rescalings
\begin{equation}
x^M \rightarrow \frac{x^M}{m}, \ \ , \ \ M=0,1,2,3,4 \ \ , \ \ 
\alpha \rightarrow \frac{\alpha}{m} \  \ , \ \ \omega \rightarrow m\omega \ \ , \ \ \phi \rightarrow \lambda \phi   \ , 
\end{equation}
where $\lambda$ is a scale. The system of equations then depends only on $\alpha$, $\omega$ and
$\kappa:=8\pi G_5 \lambda^2$ and additionally on the rescaled self-coupling $(\lambda/\eta)^{2n}$ in the non-minimal case. We then
choose $\lambda$ as follows
\begin{enumerate}
 \item for minimal boson stars $\lambda=\frac{1}{\sqrt{8\pi G_5}}$ such that $\kappa=1$\ , 
 \item for self-interacting boson stars $\lambda=\eta$ such that $\kappa=8\pi G_5 \eta^2$.
\end{enumerate}

\subsection{Interplay between rotation and Gauss-Bonnet interaction}
In order to understand how the interplay between rotation and the Gauss-Bonnet term effects the properties of boson stars, let us first remind the reader of the qualitative differences of non-rotating boson stars in
Einstein and Gauss-Bonnet gravity, respectively. 
Non-rotating boson stars in 5-dimensional Einstein gravity were first studied in \cite{Hartmann:2012gw}. These share many features with
the 4-dimensional counterparts: they exist down to a minimal frequency $\omega_{\rm min} > 0$ from where several new branches of (unstable)
solutions appear that typically form a spiralling behaviour. The critical
solution at $\omega_{\rm cr} > \omega_{\rm min}$ corresponds to a
solution that has the central value of the scalar field function, $\phi(0)$, tend to infinity. In contrast to that Gauss-Bonnet boson stars
do not have a minimal value of $\omega$. The spiral present in the former
case unfolds and solutions exist down to $\omega=0$ if $\alpha$ is large enough
and hence the GB term dominates the gravitational interaction. However, the value of $\phi(0)$ is now restricted and stays below a critical value $\phi(0)_{\rm cr} < \infty$ \cite{Hartmann:2013tca}. 
This seems to suggest that while for Einstein gravity we can localize the
scalar field and hence the energy density arbitrarily close to $r=0$,
this is impossible for the Gauss-Bonnet case. We believe that this
is related to the fact that Gauss-Bonnet gravity has a fundamental
minimal length scale inherent in it, namely the Planck length $l_{\rm Pl,5}$ which is related to the parameter $\alpha$. 

Rotating boson stars in Einstein gravity show an analogue behaviour as compared to the
non-rotating solutions \cite{Hartmann:2010pm,Brihaye:2013zha}. In the case of rotating solutions, $\phi(0)=0$ and hence the derivative
of $\phi(r)$ at zero, $\phi'(0)$, can be used as a parameter. Again, several branches
exist and the solutions exist in a limited range of $\omega$ with
$\omega_{\rm min} > 0$, while solutions can be constructed for arbitrarily large values 
of $\phi'(0)$. Now, it has been observed in \cite{Brihaye:2013zha} that this is different for the Gauss-Bonnet case.
In this latter case the value of $\phi'(0)$ is limited to a finite critical
value at which the solutions cease to exist as was pointed out in \cite{Henderson:2014dwa}. 
Here we demonstrate that the criticality of the solutions appears already
on the level of the behaviour of the functions at the origin. For that we expand the
functions close to $r=0$ taking the boundary conditions into account:
\be
  f(r\ll 1) = 1 + F_2 r^2 + O(r^4) \ , \ b(r\ll 1)= B_0 + B_2 r^2 + O(r^4) \ , \ h(r\ll 1) = r^2(1 + H_2 r^2 + O(r^4)) \ , 
\ee
\be  
  W(r \ll 1) = W_0 + W_2 r^2 + O(r^4) \ , \ \phi(r\ll 1) = \phi_1 r + O(r^3) \ ,
\ee 
where $F_2,B_0,B_2, W_0, W_2, \phi_1$ are constants to be determined numerically.
Note, however, that the equations of motion lead to several relations between these constants, namely
\be
 B_2 = B_0  \frac{\alpha \phi_1^2 + 3 (F_2 + H_2) }{3 \alpha(F_2+H_2)-3}  \ , 
 \  W_2 = - \frac{\kappa \phi_1^2 (W_0 + \omega)}{6 \alpha(F_2 + 3 H_2)- 6}    \ 
\ee
as well as
\be
\label{constraint}
3 \alpha(F_2^2 + 2 F_2 H_2 + 5 H_2^2) - 6(F_2 + H_2) - 2 \kappa \phi_1^2  = 0 \ \ .
\ee
This latter equation can be solved for either $H_2$ and $F_2$. The solutions for $H_2$ is
\begin{equation}
H_2= -\frac{F_2}{5} + \frac{1}{5\alpha} \pm \frac{\sqrt{3}}{15\alpha} \sqrt{\Delta} 
\end{equation}
with discriminant $\Delta$ given as follows
\begin{equation}
\label{deltas}
\Delta = -12 \alpha^2 F_2^2 + 24 \alpha F_2 + 10 \alpha \kappa \phi_1^2 + 3   \ \ .
\end{equation}

\begin{figure}[H]
\begin{center}
{\includegraphics[width=10cm]{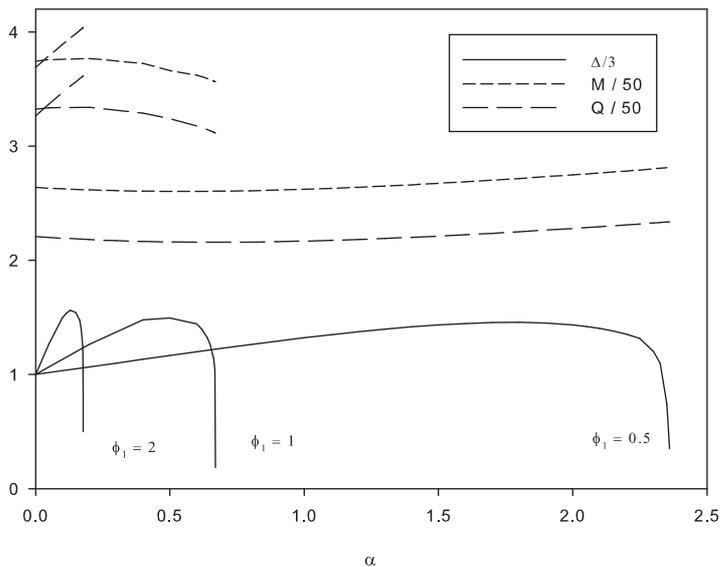}}
\end{center}
\caption{We show the quantity $\Delta/3$ (solid, see (\ref{deltas})) for different values of $\phi_1$ in dependence on $\alpha$ for rotating minimal boson stars. We also give the corresponding mass $M$ (short-dashed) and the charge $Q$ (long-dashed).
\label{delta_1_critical}}
\end{figure} 

In Fig. \ref{delta_1_critical} we show the value of $\Delta/3$ in dependence on $\alpha$ for different values of $\phi_1$. We also give the corresponding mass $M$ and the charge $Q$ of the solution. As indicated by (\ref{deltas}) we find that the larger $\phi_1$ the smaller is the value of $\alpha$ at which $\Delta=0$. This is related to the fact that
at $r=0$ the energy-mass density is dominated by the kinetic term encoded in $\phi_1$ and
hence, when $\alpha$ becomes large, Gauss-Bonnet gravity dominates the gravitational
interaction. And since Gauss-Bonnet gravity has a natural minimal length scale encoded in it, we would expect solutions ceasing to exist for sufficiently large $\phi_1$. 

From the data in Fig. \ref{delta_1_critical} we find the critical 
values of $\alpha$ and corresponding values of $M$, $Q$ and $\omega$ as 
given in Table \ref{table1}. The smaller we choose
$\phi_1$ the larger is the value of $\alpha$ at which $\Delta\rightarrow 0$.
Our data suggests (which makes also sense for dimensional reasons) that
\begin{equation}
 \alpha_{\rm cr}\approx \frac{2}{3\phi_1^2}  \ .
\end{equation}

\begin{table}
\begin{center}
\begin{tabular}{|c|c|c|c|c|}
\hline
$\phi_1$ & $\alpha_{\rm cr}$ & $M$ & $Q$ & $\omega$\\
\hline
 0.5   &  2.36              &    140.0   &   116.3 &  0.93\\
1.0 &   0.67               &      178.1    &  155.7  & 0.99\\
2.0    &  0.18            &     201.8    &    180.7 &  0.97 \\
\hline
\end{tabular}
\caption{Value of critical $\alpha$ for given value 
of $\phi_1$ and corresponding values of $M$, $Q$ and $\omega$. \label{table1}}
\end{center}
\end{table}

\begin{figure}[H]
\begin{center}
{\includegraphics[width=10cm]{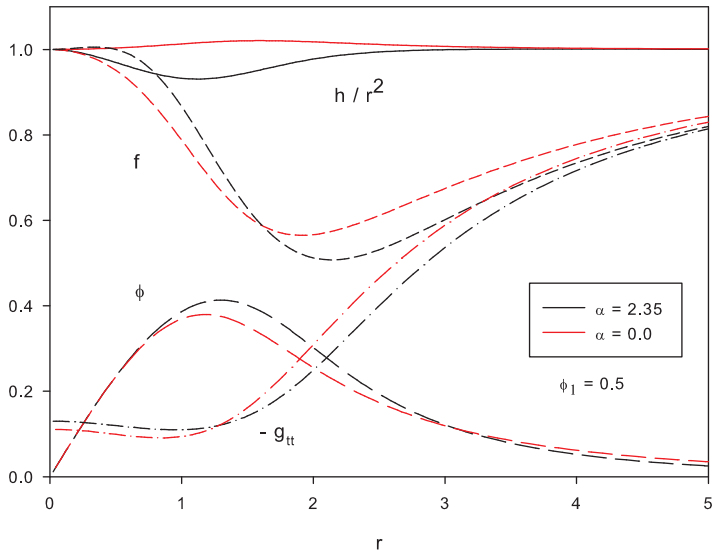}}
\end{center}
\caption{We show the metric functions $h(r)/r^2$ (solid) and $f(r)$ (short-dashed) as well as the $tt$-component
of the metric, $-g_{tt}$ (dotted-dashed) and the scalar field function $\phi(r)$ (long-dashed), for $\phi_1=0.5$ and $\alpha=0$ (red) and $\alpha=2.35$ (black), respectively.
Note that $\alpha=2.35$ is close to the critical value of $\alpha$ at 
which $\Delta\rightarrow 0$ (see also Fig. \ref{delta_1_critical} and 
Table \ref{table1} for detailed values). \label{delta_critical_pi1_05}}
\end{figure}

\begin{figure}{H}
\begin{center}
{ \includegraphics[width=10cm]{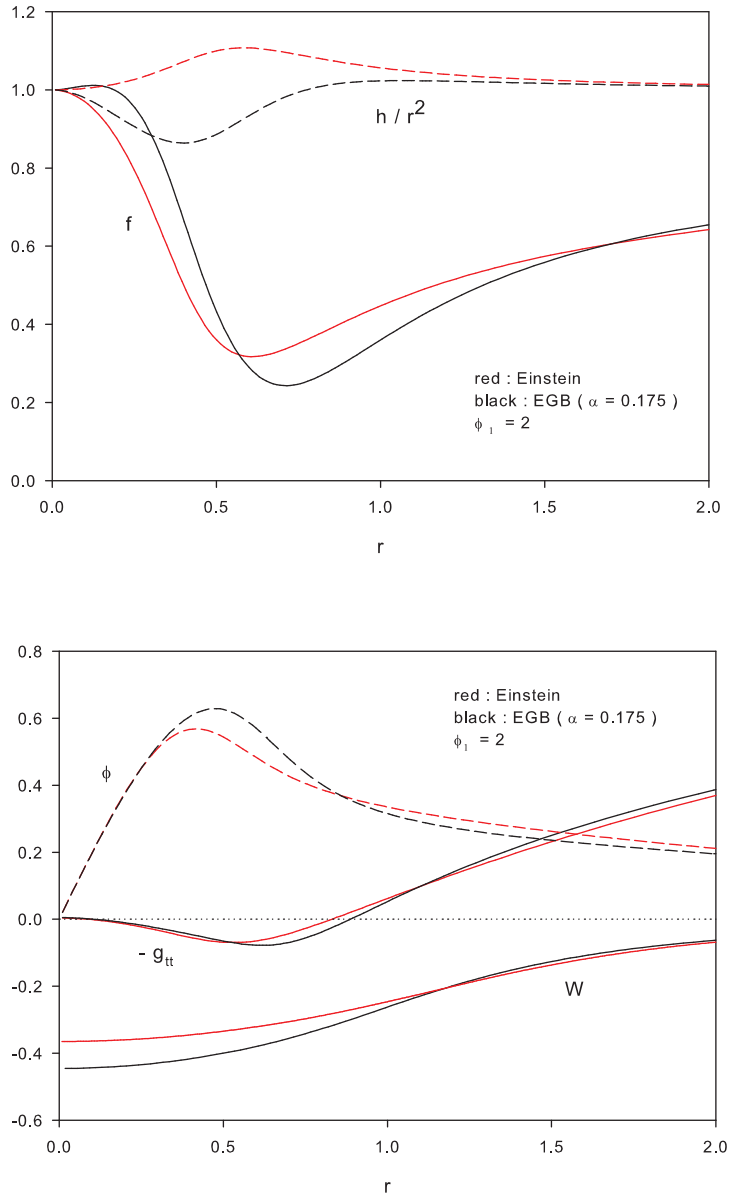}}
\end{center}
\caption{We show the metric functions $h(r)/r^2$ (short-dashed, upper plot) and $f(r)$ (solid, upper plot) as well as the $tt$-component of the metric, 
$-g_{tt}$ (solid, lower plot), the scalar field function $\phi(r)$ (short-dashed, lower plot) and the metric function $W(r)$ (solid, lower plot) for $\phi_1=2$ and $\alpha=0$ (red) and $\alpha=0.175$ (black), respectively. Note that
$\alpha=0.175$ is close to the critical value of $\alpha$ at which $\Delta\rightarrow 0$ 
(see also Fig. \ref{delta_1_critical} and 
Table \ref{table1} for detailed values).
\label{fig3}}
\end{figure}

In Fig. \ref{delta_critical_pi1_05} we show the behaviour of the metric functions
for $\phi_1=0.5$ and $\alpha=0$ as well as $\alpha=2.35\approx \alpha_{\rm cr}$.
As is clearly seen from this figure, the metric functions $f(r)$ and $h(r)$ change their
behaviour close to $r=0$. While for $\alpha=0$ the metric function $f(r)$ decreases
from unity close to the origin, it increases in the case of strong Gauss-Bonnet coupling.
The metric function $h(r)/r^2$ shows the opposite behaviour: for $\alpha=0$ it increases
from unity, while in the Gauss-Bonnet case it decreases. This can also be seen in 
Fig. \ref{fig3}, where we show the metric functions as well as the scalar field
function $\phi(r)$ for $\phi_1=2$ and $\alpha=0$ as well as $\alpha=0.175\approx \alpha_{\rm cr}$. While the qualitative behaviour of $W(r)$, $g_{tt}$ and $\phi(r)$ close
to the origin does not change, the metric functions $f(r)$ and $h(r)/r^2$ show 
qualitatively different behaviour. This is a clear indication of the fact that for large
enough $\alpha$ the Gauss-Bonnet interaction dominates the gravitational interaction
and hence the scalar curvature is no longer given by the energy-momentum content
of the space-time as in Einstein gravity where $R\sim T$. This is clearly seen in Fig. \ref{compare_R_T}, where we compare the
local scalar curvature $R$ with $-2T/3$ for $\alpha=0$ and $\alpha=0.175$ close to the maximal possible value of $\alpha$, respectively, for
$\phi_1=2$. For the former case we know from the Einstein equation that $R=-2T/3$ (letting $8\pi G_5\equiv \kappa=1.0$) and we confirmed
this equality numerically (and hence checked our numerical procedure to be valid). For $\alpha=0.175$ the scalar curvature $R$ and $-2T/3$ differ close to the origin. While the local maximum of 
the trace of the energy-momentum tensor, $T$, is located roughly at the maximum of the scalar field function $\phi(r)$ 
(see Fig. \ref{delta_critical_pi1_2}) and the increase of the Gauss-Bonnet interactions increases the $r$ at which the maximal
energy-momentum content is located, the scalar curvature $R$ at $r=0$ decreases with increasing $\alpha$ and shows no
longer a pronounced minimum at some finite value of $r$. 

\begin{figure}[H]
\begin{center}
{ \includegraphics[width=10cm]{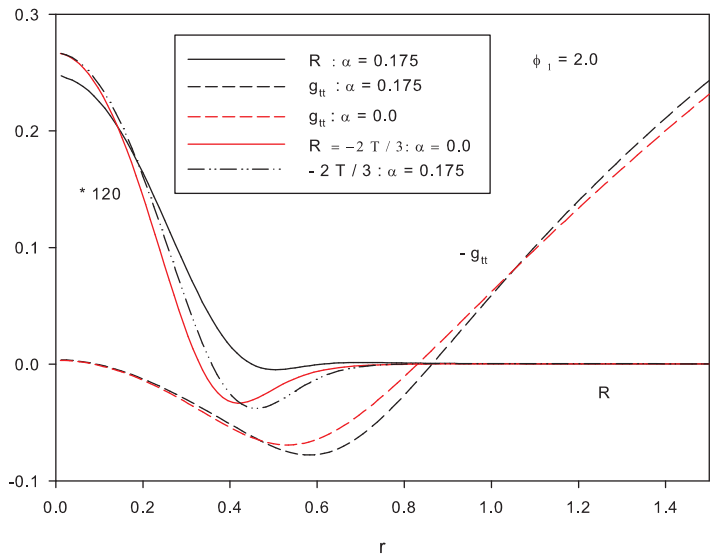}}
\end{center}
\caption{We show the $tt$-component of the metric, $-g_{tt}$ (dashed), as well as the Ricci scalar $R$ (solid)
for $\alpha=0$ (red) and $\alpha=0.175$ (black), which is close to the maximal possible value of $\alpha$ for which $\Delta\rightarrow 0$ for the chosen
value of $\phi_1=2$
(see also Fig. \ref{delta_1_critical}). For $\alpha=0$ it follows from the Einstein equation that $R=-2T/3$ (with our rescalings such that
$\kappa=1$), while we also show $-2T/3$ (black dotted-dashed) for $\alpha=0.175$ to demonstrate that the local energy-momentum
no longer determines the local scalar curvature in the case where the Gauss-Bonnet term dominates the gravitational interaction.
\label{delta_critical_pi1_2}
\label{compare_R_T} }
\end{figure}

\subsection{Instability of minimal boson stars in 5 dimensions} 
The biggest fraction of the results presented in this subsection correspond 
to that of a massive scalar field
without self-interaction. This case -- at least to our knowledge -- has not be studied in detail in the literature
so far. As pointed out above, we can rescale coordinates and fields such that
we can set $\kappa=m=1$ in this case, such that the only free parameter is the angular
frequency $\omega$ in Einstein gravity. For Gauss-Bonnet gravity $\alpha$ is a second free parameter.

\begin{figure}[H]
\begin{center}
\subfigure[$M$, $Q$]
{\label{for_terence}\includegraphics[width=8cm]{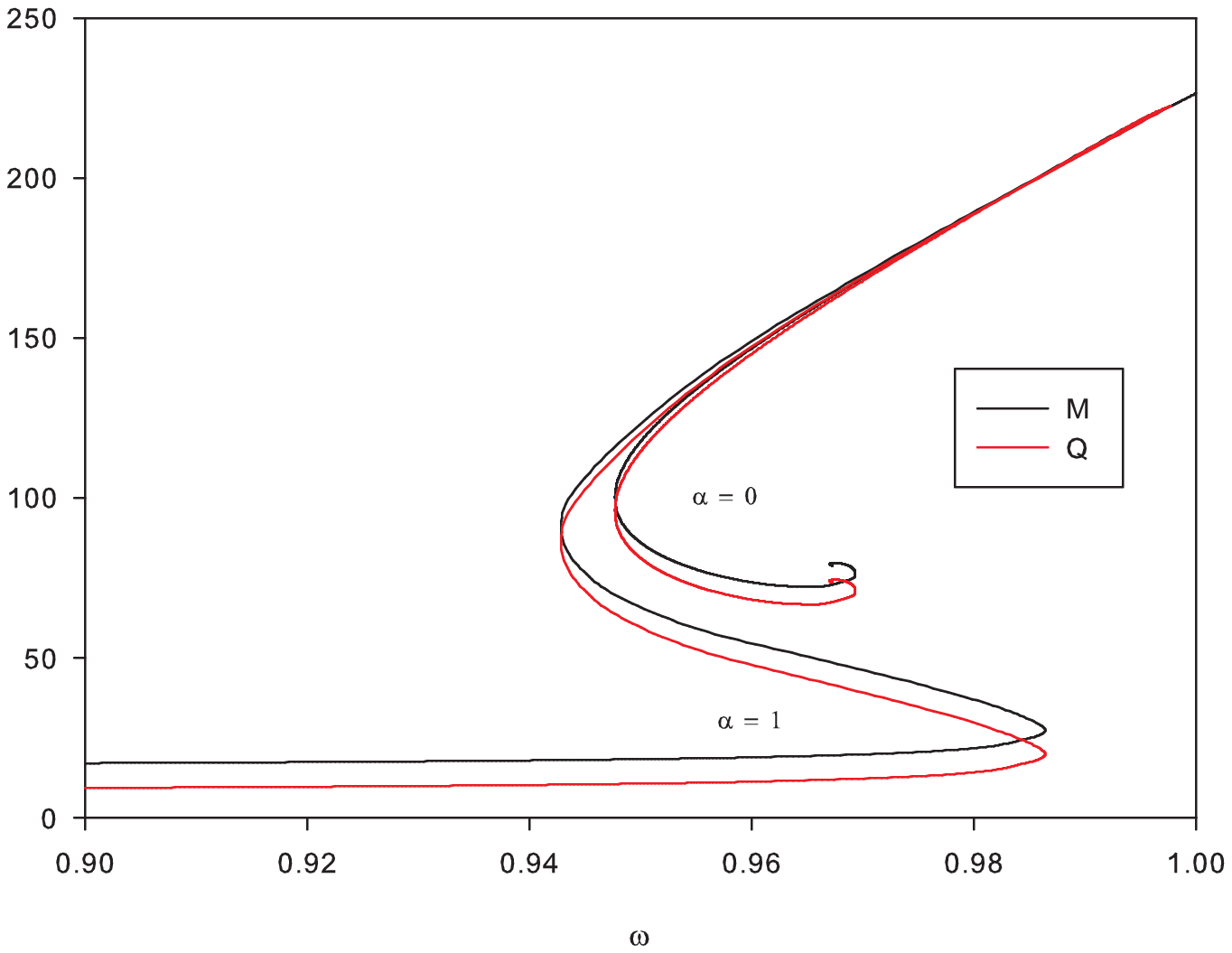}}
\subfigure[$\phi(0)$ , $b(0)$, $R(0)$]
{\label{hyper_r}\includegraphics[width=8cm]{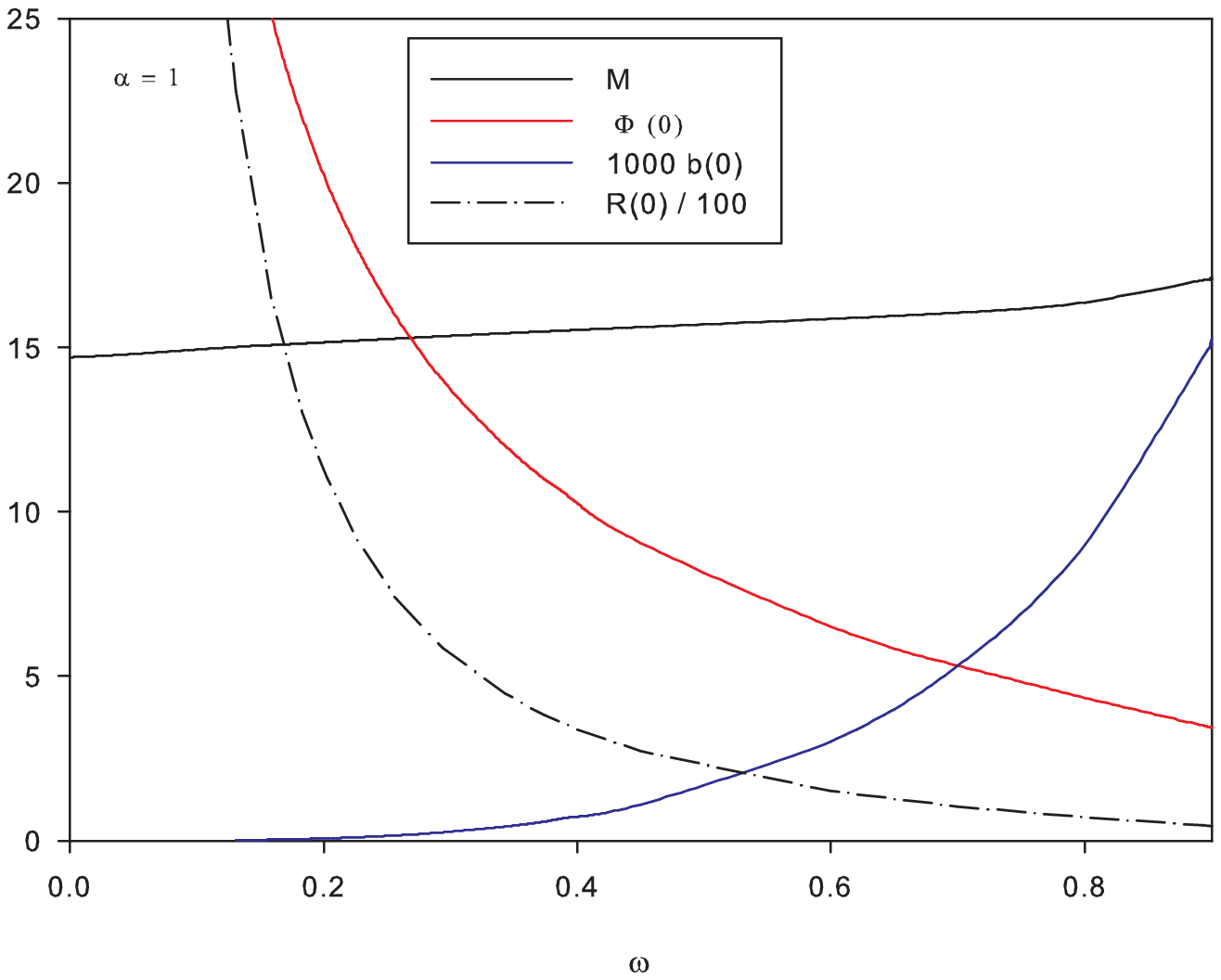}}
\end{center}
\caption{We show the mass $M$ (black) and the charge $Q$ (red) in dependence on the frequency $\omega\in [0.9:1]$ 
for non-rotating minimal boson stars and for $\alpha=0$ and $\alpha = 1$, respectively (left).
We also show the dependence of the value of the scalar field function $\phi(r)$ at the origin, $\phi(0)$ (solid, red), of the value of 
the metric function $b(r)$ at the origin, $b(0)$ (solid, blue), of the value of the scalar curvature $R$ at the origin, $R(0)$ (dotted-dashed, black), as well as of the mass
$M$ (solid, black)
on $\omega\in \ ]0:0.9]$ for $\alpha = 1$ (right). 
\label{mass_w_m_q}
}
\end{figure}

\subsubsection{Non-rotating solutions}
In contrast to the rotating case, non-rotating solutions have a non-vanishing value
of the scalar field at the origin, $\phi(0)$, while the derivative vanishes there.
Hence, non-rotating boson stars are typically characterized by the value of the scalar
field at the origin, which is determined by the value of $\omega$ and vice versa. 

We first discuss the case $\alpha=0$. We find that minimal boson stars
exist for $\omega\in [0.9477:1]$. This frequency does not uniquely 
 characterize the boson stars. Indeed several solutions exist with the same value of $\omega$, however,
 with different values of $M$ and $Q$. 
This is shown in Fig. \ref{mass_w_m_q} (left), where we give the mass $M$ and the charge $Q$ 
as function of $\omega$. Several branches of solutions are present (as in the case
of self-interacting boson stars) and the typical spiraling behaviour is seen.

 \begin{figure}[H]
\begin{center}
\subfigure[$M$, $Q$, $\phi(0)$, $b(0)$]
{\label{for_terence}\includegraphics[width=8cm]{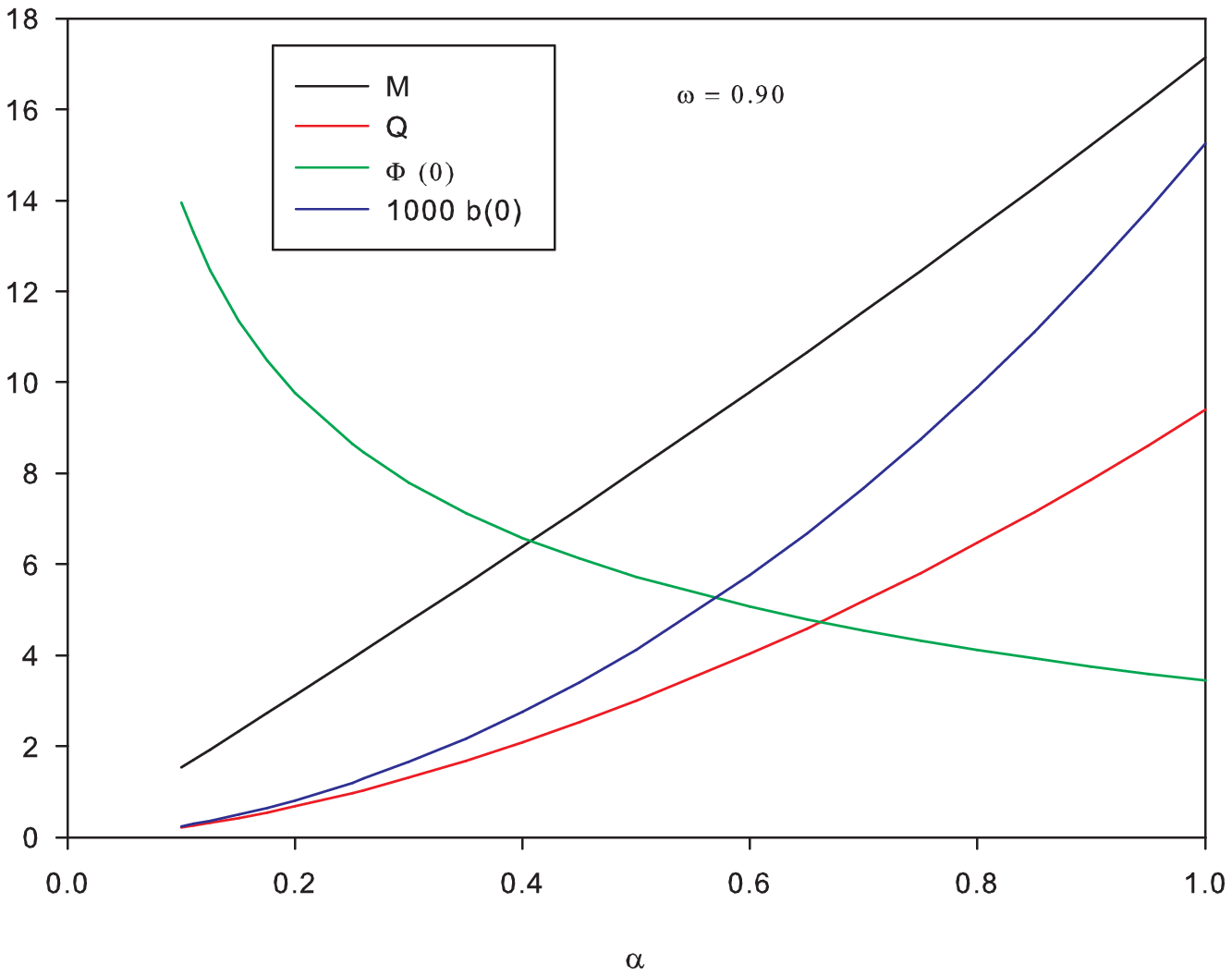}}
\subfigure[$b(0)$, $f_m$, $R(0)$]
{\label{hyper_r}\includegraphics[width=8cm]{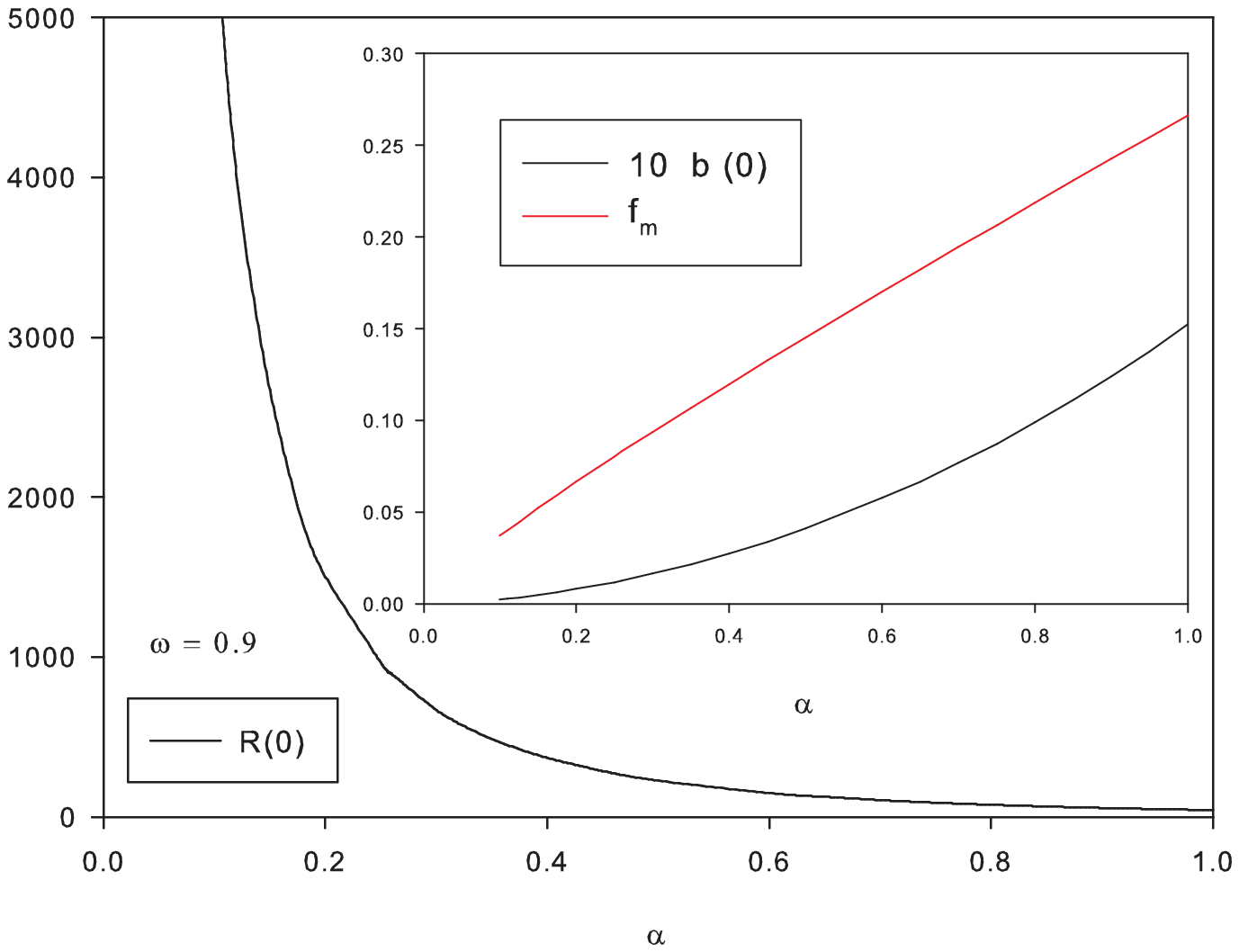}}
\end{center}
\caption{We show the mass $M$ (solid, black), the charge $Q$ (solid,red), the value of the metric function
$b(r)$ at the origin, $b(0)$ (solid, blue), as well as the value of the scalar field function $\phi(r)$ at the origin, $\phi(0)$ (solid, green), 
in dependence on $\alpha$ for non-rotating minimal boson stars with $\omega=0.9$ (left).
We also give the value of the metric function $b(r)$ at the origin, $b(0)$ (solid, subfigure), the minimal
value of the metric function $f(r)$, $f_m$ (red, subfigure), as well as the value of the scalar
curvature $R$ at the origin, $R(0)$ (black), in dependence on $\alpha$ for these solutions (right).}
\label{alpha_to_zero_data}
\end{figure}  

 The limit $\phi(0) \to 0$ corresponds to $\omega \to  1$
 and the scalar field function 
 approaches the null function uniformly. Interestingly, the mass and charge are finite
 in this limit $M_{\omega \to 1} = Q_{\omega \to 1} \approx 226.0$. This was also observed
 for boson stars with self-interaction in \cite{Hartmann:2010pm} and the presence of this
 ``mass gap'' is a special feature in 5 space-time dimensions.
 Boson stars  exist for arbitrarily large values of $\phi(0)$, however the mass, 
 the charge and $\omega$ remain finite. The spiral ends at $\phi(0)\rightarrow \infty$ with
 $\omega \sim 0.96$, $M \sim 78.95$ and $Q \sim 73.7$.
 
  The scenario is different for the Gauss-Bonnet case ($\alpha > 0$). 
  In the region $\omega \sim 1$, (i.e. for $\phi(0) \ll 1$) the solution
 is rather insensitive to $\alpha$, which is natural since the local energy-momentum
 content is small. However, quantitative and qualitative differences occur 
 when $\phi(0)$ is increased.
 As shown in Fig. \ref{mass_w_m_q} for $\alpha = 1$ the spiral has disappeared 
 and solutions now exist for all $\omega\in \ ]0:1]$.
 We also show some parameters characterizing the solution in Fig. \ref{mass_w_m_q} (right). As can be seen in this figure, 
 we find that the value of the metric function $b(r)$ at the origin, $b(0)$, tends to zero. This suggests the appearance
 of a singularity at $r=0$ in the limit $\omega \rightarrow 0$. In addition, we observe that the Ricci scalar $R$ becomes very large in this limit, while the mass stays finite.
 
 In order to understand this critical behaviour better, we have also studied the case of a fixed $\omega$ and varying $\alpha$,
 which, in fact, turned out to be more feasible numerically. Our results for $\omega=0.9$ are shown in Fig. \ref{alpha_to_zero_data}, where we show the dependence of the mass, charge and several parameters characterizing the solution on $\alpha$. As is clear from
this figure, the mass and charge decrease when decreasing $\alpha$ for fixed $\omega$.
At the same time $\phi_1$ increases. In order to understand the critical behaviour we
have also plotted the value of the metric function $b(r)$ at the origin, $b(0)$, 
as well as the minimal value of the metric function $f(r)$, $f_m$. $f_m\rightarrow 0$ would indicate 
the formation of an extremal black hole which would carry ``scalar hair''.
However, as was shown in \cite{Brihaye:2012cb} the near-horizon $AdS_2\times S^3$ 
geometry of an extremal Gauss-Bonnet black hole does not support scalar hair. Hence
the limiting solution is a singular solution with $b(0)\rightarrow 0$.

From the figures presented above it is obvious that all the solutions without self-interaction
have $M > Q$ which indicates that these solutions are always classically unstable. This contrasts with the 4-dimensional
case, where it was shown that minimal boson stars are classically stable \cite{Lee:1988av}. Following the arguments
in this latter paper it is easy to show why this should be the case. Consider the boson star made out of $Q$ scalar quanta
of mass $m$. These scalar quanta have each kinetic energy $\sim p \sim \lambda^{-1} \sim R^{-1}$, where
$p$ is the linear momentum of the scalar quantum, $\lambda$ its average wavelength and $R$ the radius of the boson star.
The kinetic energy of the boson star made out of $Q$ quanta is hence $E_k\sim Q/R\sim M/(mR)$, where we assume that
$M\sim mQ$. In 4 dimensions the gravitational energy is $E_{\rm g,4}\sim -G_4 M^2/R$, where $G_4$ is the 4-dimensional Newton's constant.
It is thus possible to find an equilibrium between the kinetic energy and the gravitational energy. As pointed out in
\cite{Lee:1988av}, the kinetic energy makes the star expand until the gravitational energy dominates the system and -- because of its
attraction -- allows to have a stable star. Now, this is different in 5 dimensions, where the gravitational energy
is $E_{\rm g,5}\sim -G_5 M^2/R^2$. At small $R$ the gravitational energy will always dominate and hence stable
configurations are not possible. This changes when one includes a repulsive self-interaction which can balance
the gravitational attraction and allow for stable boson stars. We show the ratio $M/Q$ in Fig. \ref{MQ_compare} for non-rotating
boson stars without self-interaction and self-interaction potential (\ref{pot_self}), respectively. It is clear that
without self-interaction the ratio $M/Q$ is always larger than unity and hence the boson stars are always classically
unstable. We also demonstrate that the value of the Gauss-Bonnet coupling does not change anything as far as this conclusion is
concerned. Now, this changes when the self-interaction is present. For $\kappa$ small stable solutions exist, while
for sufficiently large $\kappa$ the solutions become again unstable. This latter observation is related to the fact that
for $\kappa$ too large the gravitational energy dominates the kinetic energy (as argued above) and hence stable configurations
are not possible. The figure demonstrates also that the inclusion of the Gauss-Bonnet interaction does not change much for the
self-interacting case either. 

\begin{figure}[H]
\begin{center}
{\includegraphics[width=10cm]{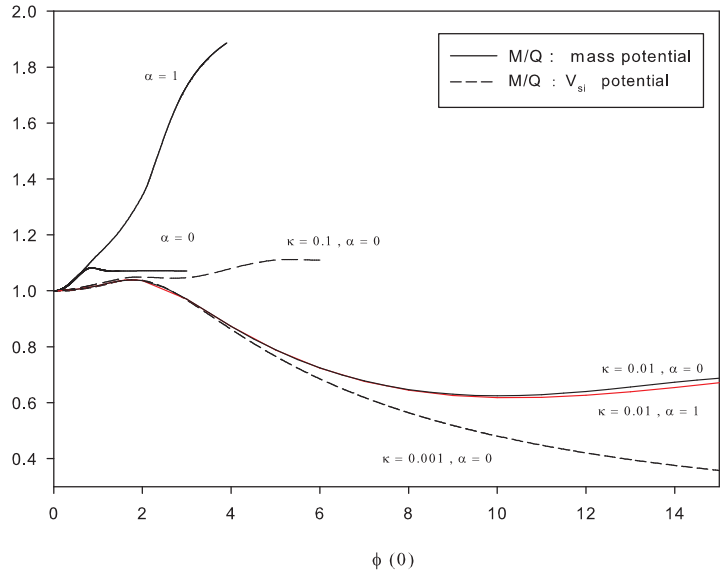}}
\end{center}
\caption{We show the ratio of the mass $M$ and the charge $Q$, $M/Q$ in dependence on the value of the scalar field function $\phi(r)$ at the
origin, $\phi(0)$, 
for non-rotating boson stars with $\alpha=0$ and $\alpha=1$, respectively. 
We compare the case of boson stars without self-interaction (solid) and that with self-interaction potential $V_{\rm si}$ (dashed).
Note that in the case without self-interaction we can scale $\kappa\equiv 1$ without loss of generality, while
we have adapted the re-scaling $\phi\rightarrow \eta\phi$ for the case with self-interaction and hence study different values of $\kappa$.
\label{MQ_compare}
}
\end{figure}

 \begin{figure}[H]
\begin{center}
{\includegraphics[width=10cm]{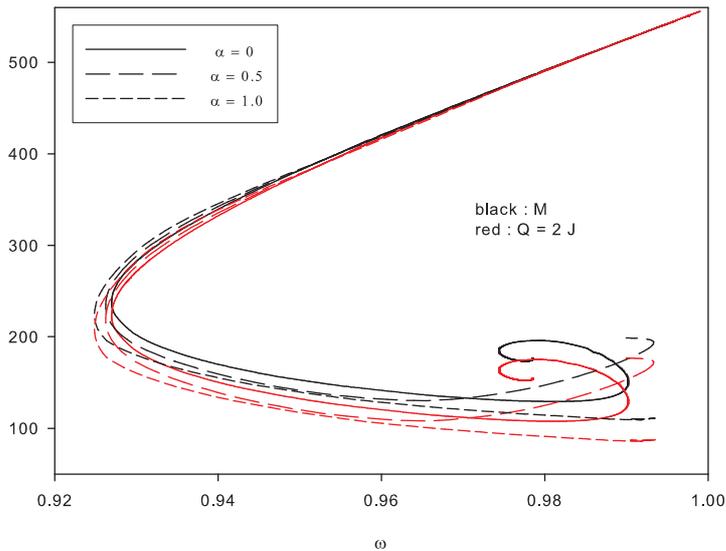}}
\end{center}
\caption{We show the mass $M$ (black) and the charge $Q$ (red) in dependence on the frequency $\omega$ 
for rotating minimal boson stars and for $\alpha=0$ (solid), $\alpha = 0.5$ (long-dashed) and $\alpha=1.0$ (short-dashed), respectively.
Note that the angular momentum is related to the Noether charge like $J=Q/2$.
\label{rotating_non_interacting}
}
\end{figure}

\subsubsection{Rotating solutions}
In Fig. \ref{rotating_non_interacting} we show the mass $M$ and charge $Q$ for rotating minimal boson stars 
and different values of $\alpha$. The qualitative pattern is very similar to that of non-rotating solutions.
For $\alpha=0$ we find the typical spiraling behaviour, while for large values of $\alpha$ the solutions cease to exist
at some critical value of $\phi_1$, respectively $\omega$. 

As far as the classical stability of rotating minimal boson stars the conclusions are very similar to the non-rotating case. As can be seen from Fig. \ref{rotating_non_interacting}
we find that independent of the choice of $\alpha$ the solutions have $M > Q$ and are hence classically unstable. While we would expect
a repulsive centrifugal force to help balance the gravitational attraction in this case, this is not sufficient to render the solutions
stable. 

\subsection{Ergoregions}

\begin{table}[h]
\begin{center}
\begin{tabular}{|c|c|c|c|c|}
\hline
$\phi_1$ & $\alpha$ & $r_{\rm i}$ & $r_{\rm o}$ & $V_{\rm ergo}$ \\
\hline
 0.5 &   0.00             &    n.a.    &   n.a. & n.a.\\
 0.5 &   2.36   &    n.a.     & n.a. & n.a. \\
 1.0 &   0.00             &    0.41    &   0.92 & 6.43  \\
 1.0 &   0.67   &    0.61     & 0.98 & 6.48 \\
 2.0    &  0.00          &   0.09     &  0.82 & 6.10  \\
 2.0 &   0.18   &   0.09      &  0.86 & 6.30\\
\hline
\end{tabular}
\caption{Value of the inner and outer radius of the ergoregion, $r_i$ and $r_o$, 
respectively as well as the proper volume of the ergoregion $V_{\rm ergo}$ for different values of $\phi_1$ and $\alpha$. Note that the
non-vanishing values of $\alpha$ are close to $\alpha_{\rm cr}$ at which $\Delta\rightarrow 0$ for the respective value of $\phi_1$.
The abbreviation ``n.a.'' implies that the space-time does not possess an ergoregion. \label{table2}}
\end{center}
\end{table}

\begin{figure}[H]
\begin{center}
{\includegraphics[width=10cm]{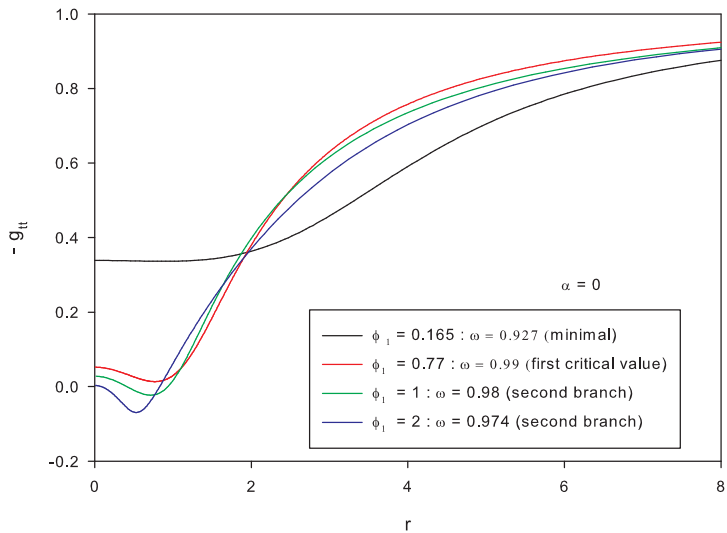}}
\end{center}
\caption{We show the $tt$-component of the metric for different values of $\phi_1$ and $\omega$, respectively,
and for minimal boson stars with $\alpha=0$. \label{ergoregion_functions} }
\end{figure}

Rotating black holes typically possess ergoregions between the static limit surface and the event horizon. This is the region in
which the asymptotic time-like Killing vector field becomes space-like and processes such as the energy extraction from black holes
(Penrose process) become possible. Rotating objects without event horizon can also possess such ergoregions under specific conditions.

We find that these ergoregions also exist for our solutions. This can be seen in Fig. \ref{delta_critical_pi1_05} for $\phi_1=0.5$    
and Fig. \ref{compare_R_T} for $\phi_1=2$, respectively. As is obvious from this latter figure, the value of $\alpha$
does not influence the ergoregion much. It shifts the radius of the ergosphere to slightly larger values of $r$, but
otherwise does not influence the qualitative shape. However, it is the value of $\phi_1$ that triggers the existence
of an ergoregion. For $\phi_1=0.5$ no ergoregion exists, while for $\phi_1=1.0$ it is present already and persists to be present
for $\phi_1=2$. We have investigated this in more detail.
In Fig. \ref{ergoregion_functions} we show the $tt$ component of the metric for different
values of $\phi_1$ and $\omega$, respectively. We find that when increasing $\phi_1$ an ergoregion appears for sufficiently
large $\phi_1$. For $\alpha=0$ this happens on the second branch of solutions. 

In Table \ref{table2} we give the value of the inner and outer radius of the
ergoregion denoted by $r_i$ and $r_o$, respectively, for two different values of $\phi_1$ and for the Einstein limit ($\alpha=0$) and the maximal possible value of $\alpha$, $\alpha_{\rm cr}$, at which $\Delta\rightarrow 0$. For both values of $\phi_1$ we observe that the value of $r_o$ is slightly increased. For $\phi_1=1.0$ the inner radius is
pushed outwards close to the critical value of $\alpha$, which is also the maximal
possible value. Here, the GB interaction is strongest and its influence gets stronger when
approaching the origin. For $\phi_1=2$ the maximal possible $\alpha$ is
relatively small, hence we would expect the GB interaction to play little influence here.
For a fixed value of $\alpha$ we observe that  an increase of $\phi_1$ leads to
two things: (a) a shift of the inner and outer radius to smaller values of $r$ and
(b) the increase of the extend of the ergoregion in $r$. 
When computing the proper volume of the ergoregion $V_{\rm ergo}=\int\limits_{r_i}^{r_o} \sqrt{-g_4} {\rm d}^4 x$, where $g_4$ is the determinant
of the spatial part of the metric, we notice that this increases slightly with $\alpha$ and decreases
with increasing $\phi_1$. For $\alpha=0$ we find that the ergoregion first appears at $\phi_1\approx 0.842$ when increasing $\phi_1$ from zero.
The volume $V_{\rm ergo}$ then shows a sharp increase from zero  and stays around the value $6$ for $\phi_1 \geq 1.0$.

\section{Summary and Outlook}
In this paper we have focused our studies on minimal boson stars in 5 space-time dimensions. We have pointed out that these minimal boson stars are always classically
unstable -- in contrast to their 4 dimensional counterparts. This is due to the fact that the gravitational interaction always dominates at small $r$ and hence a stable
configuration is not possible. Moreover, the boson stars possess ergoregions for
sufficiently large increase of the scalar field function at the origin.
We observe that the Gauss-Bonnet interaction alters the location and appearance of
this ergoregion only marginally. This is also related to the fact that rotating Gauss-Bonnet boson stars
exist only up to a maximal value of the Gauss-Bonnet coupling and hence solutions with arbitrarily large Gauss-Bonnet
interaction do not exist. These ergoregions make the solutions suffer (additionally) from an {\it ergoregion instability}.  

It would also be interesting to check whether minimal boson stars in dimensions higher than 5 are classically unstable.
We believe they are because the power of the fall-off of the gravitational potential increases with the number of dimensions, but leave this
for future work.

In addition, one can construct boson stars with non-equal angular momenta and check whether the solutions possess
ergoregions. These ergoregions -- if they exist -- will be different in shape to the ones discussed here.
The ergoregions of rotating boson stars with equal angular momenta are spherical hypershells, while those of non-equal angular
momenta solutions will have an angular dependence. It would be very interesting to see whether the ergoregion instability is
present and what effects it has. \\
\vspace{1cm}

{\bf Acknowledgments} YB would like to thank the Belgian FNRS for financial support.

\end{document}